\documentclass{aa}  

\usepackage{graphicx}
 \usepackage{lscape}
\usepackage{graphicx}
\usepackage{color}
\usepackage{amsmath}

\usepackage{natbib}
\usepackage{float}

\usepackage{txfonts}
\begin{document}

   \title{A study of the C$_3$H$_2$ isomers and isotopologues: first interstellar detection of HDCCC \thanks{Based on observations carried out with the IRAM 30m Telescope. IRAM is supported by INSU/CNRS (France), MPG (Germany) and IGN (Spain)}}


   \author{S. Spezzano
          \inst{1,3}
          \and
          H. Gupta\inst{2}
              \and
          S. Br\"unken\inst{3}    
          \and
          C. A. Gottlieb\inst{4}    
          \and
          P. Caselli\inst{1}    
          \and
          K. M. Menten\inst{5}    
          \and
         H. S. P. M\"uller\inst{3}    
          \and
          L. Bizzocchi\inst{1}   
           \and
          P. Schilke\inst{3}
              \and
          M. C. McCarthy\inst{4}
               \and
          S. Schlemmer\inst{3}         }

   \institute{Max-Planck-Institut f\"ur extraterrestrische Physik, Giessenbachstr. 1, 85748 Garching, Germany
     \and
   California Institute of Technology, 770 S. Wilson Ave., M/C 100-22, Pasadena, CA 91125 (current address: Division of Astronomical Sciences
National Science Foundation,4201 Wilson Boulevard, Suite 1045, Arlington, VA 22230)
   \and
   I. Physikalisches Institut, Universit\"at zu K\"oln, Z\"ulpicher Str. 77, 50937 K\"oln, Germany
   \and
   Harvard-Smithsonian Center for Astrophysics, Cambridge, MA 02138, and School of Engineering \& Applied Sciences, Harvard University, Cambridge, MA 02138, USA
   \and
   Max-Planck Institut f\"ur Radioastronomie, Auf dem H\"ugel 69, 53121 Bonn, Germany
 }


 
  \abstract
{The partially deuterated linear isomer HDCCC of the ubiquitous cyclic carbene ($c$-C$_3$H$_2$) was observed in the 
starless cores TMC-1C and L1544 at 96.9~GHz, and a confirming line was observed in TMC-1 at 19.38~GHz. 
To aid the identification in these narrow line sources,
four centimetre-wave rotational transitions (two in the previously reported $K_a =0$  ladder, and two new ones in the $K_a =1$ ladder),
and 23 transitions in the millimetre band between 96 and 272~GHz were measured in high-resolution laboratory spectra.
Nine spectroscopic constants in a standard asymmetric top Hamiltonian allow the principal transitions of astronomical interest in the $K_a \le 3$ rotational ladders to be calculated to within 0.1~km~s$^{-1}$ in radial velocity up to 400~GHz.
Conclusive evidence for the  identification of the two astronomical  lines of HDCCC was provided by the $V_{\rm{LSR}}$ which is the same 
as that of the normal isotopic species (H$_2$CCC) in the three narrow line sources.
In these sources, deuterium fractionation in singly substituted H$_2$CCC (HDCCC/H$_2$CCC $\sim4\%\text{-}19\%$) is comparable to that in $c$-C$_3$H$_2$ ($c$-C$_3$H$_2$/$c$-C$_3$HD $\sim5\%\text{-}17\%$), and similarly in doubly deuterated 
$c$-C$_3$H$_2$ ($c$-C$_3$D$_2$/$c$-C$_3$HD $\sim3\%\text{-}17\%$), implying that the efficiency of the deuteration processes in the
H$_2$CCC and $c$-C$_3$H$_2$ isomers are comparable in dark clouds.}

   \maketitle
%

\section{Introduction}

The study of deuterated molecules in the radio band yields constraints on the physical and chemical properties of the interstellar gas in
the early stages of low-mass star formation \citep{cas12}.
Cyclopropenylidene ($c$-C$_3$H$_2$)
is one of the molecules frequently used in studies of deuterium fractionation in cold dark clouds, because 
lines of its rare isotopic species are relatively intense, the degree of fractionation is high, 
and $c$-C$_3$H$_2$ is believed to form solely by gas phase reactions \citep{ger87,bel88,tal01,tur01,lis12,spe13}.
During recent observations of the doubly deuterated species $c$-C$_3$D$_2$ toward the starless cores TMC-1C and L1544, 
\citet{spe13} observed an unidentified line in the millimetre band that was tentatively assigned to the singly deuterated form of propadienylidene (H$_2$CCC),\footnote{Propadienylidene  [Chemical Abstracts Services CAS \#: 60731-10-4] is usually designated in the spectroscopic and astronomical literature as H$_2$CCC \citep[e.g.,][]{vrt90,cer91}, to distinguish it from its cyclic isomer $c$-C$_3$H$_2$, and because it has a same structure and symmetry as the well known molecules formaldehyde (H$_2$CO) and ketene (H$_2$CCO).   Although propadienylidene is occasionally  designated as $l$-C$_3$H$_2$, in the work here we follow the example of spectroscopists and astronomers and refer to propadienylidene as H$_2$CCC.} a highly polar metastable carbene ($\mu = 4.17$~D) about 14~kcal~mol$^{-1}$ (7045~K) higher in energy than the more stable isomer $c$-C$_3$H$_2$  \citep{wu10}.

Propadienylidene (H$_2$CCC) was detected in space by \citet{cer91} and \citet{kaw91} shortly after its millimetre-wave rotational spectrum was measured in the laboratory  \citep{vrt90}. Initially, it was observed toward the cold dark cloud TMC-1 in the centimetre and millimetre bands, and the carbon-rich
asymptotic giant branch (AGB) star IRC+10216 in the millimetre band.
Here, we report the detection of a line of HDCCC in the prestellar cores TMC-1C and L1544 in the millimetre band at 96.9~GHz, and a confirming line in TMC-1 in the centimetre band at 19.38~GHz.

Although H$_2$CCC was first observed nearly 25~years ago in the laboratory and the interstellar gas, the deuterated species had not been observed in any 
astronomical source and the rotational spectrum of the partially deuterated species HDCCC had not been measured in the millimetre band prior to this work,
but the two lowest transitions in the centimetre band had been measured earlier \citep{kim05}. In support of the astronomical identification of HDCCC, and future studies of the C$_3$H$_2$ isomeric system in the interstellar gas, 
the principal millimetre-wave rotational transitions of astronomical interest have now been measured in high resolution laboratory spectra.

\section{Laboratory and Astronomical Observations}

\subsection{Laboratory}
\label{sec:Laboratory}
Rotational lines of HDCCC were observed with the same 3~m long free space double-pass absorption spectrometer 
used to measure the millimetre-wave spectrum of $c$-C$_3$D$_2$ \citep{spe12}. Guided by frequencies calculated with an effective rotational and centrifugal distortion constant derived from the two lowest transitions 
measured at centimetre wavelengths in a supersonic molecular beam \citep{kim05}, lines of HDCCC were
observed in the millimetre band.
Following optimization of the normal isotopic species (H$_2$CCC),  
the most intense lines of HDCCC were observed in a low pressure ($\approx$18~mTorr) DC discharge (140~mA) through a statistical mixture of deuterated acetylene (50\% HCCD, 25\% DCCD, and 25\% HCCH), carbon monoxide (CO), and argon (Ar) in a molar ratio of 10:5:1, with the walls of the discharge cell cooled to 150~K.  
The acetylene sample was produced in real time by dropping an equal mixture of normal (H$_2$O) and heavy water 
(D$_2$O) on calcium carbide (CaC$_2$).
Under these conditions, the signal-to-noise (S/N) of lines of HDCCC near 160~GHz was $\ge10$ in 15~minutes of integration.
In addition, the two lowest $K_a = 0$ lines in the centimetre band at 19 and 38~GHz reported earlier by \citet{kim05} were remeasured with the FT microwave spectrometer described in \citet{spe12}, and the two $K_a = 1$ transitions ($2_{1,2} - 1_{1,1}$ and $2_{1,1} - 1_{1,0}$) were also measured. 

In all, 27 rotational lines between 19 and 272~GHz with   $J \le 14$ and $K_a \le 3$ (Table~1) are
reproduced to an rms uncertainty (28~kHz) that is comparable to the measurement uncertainties with 9 spectroscopic constants in Watson's $S$ reduced Hamiltonian: three rotational constants, four fourth order distortion constants ($D_K$ was constrained to a theoretical value owing to the high correlation with $A$), 
and two sixth order distortion constants (Table~\ref{spectro_parameter}).

\begin{table}
\caption{Laboratory frequencies of  HDCCC}
\label{HDCCC_lines}
\begin{tabular}{cccr}
\hline\hline
$J'_{K_a',K_c'} - J''_{K_a'',K_c''}$      & \multicolumn{1}{c}{Frequency}   & \multicolumn{1}{c}{Unc.}  
&  \multicolumn{1}{c}{$O-C$\tablefootmark{a}}  \\
\multicolumn{1}{c}{}                              &  \multicolumn{1}{c}{(MHz)}          &  \multicolumn{1}{c}{(kHz)}  
&  \multicolumn{1}{c}{(kHz)}  \\
\hline
 $1_{0,1} - 0_{0,0}$    &  19384.5098           &  2     & $0.5$ \\
 $2_{1,2} - 1_{1,1}$    &  38283.5891           &  2     & $-0.6$ \\
 $2_{0,2} - 1_{0,1}$    &  38768.0019           &  2     & $1.0$ \\
 $2_{1,1} - 1_{1,0}$    &  39251.4274           &  2     & $-1.2$ \\
 $5_{0,5} - 4_{0,4}$    &  96902.196             & 20    &  $0.1$    \\
 $6_{0,6} - 5_{0,5}$    & 116271.438            &  20   & $-6.5$ \\
 $7_{0,7} - 6_{0,6}$    & 135634.586            &  20   & $-8.5$ \\
 $8_{0,8} - 7_{0,7}$    & 154990.661            & 20    &  $26.5$ \\
 $9_{0,9} - 8_{0,8}$    & 174338.544            &  20   &  $-13.9$ \\
$10_{0,10} - 9_{0,9}$ & 193677.324            &  20   & $-40.5$ \\
$10_{1,9}    -  9_{1,8}$    & 196204.056       &  20   & $-14.4$ \\
$12_{1,12} - 11_{1,11}$ & 229614.027       &  23   & $55.4$ \\
$12_{2,11} - 11_{2,10}$ & 232512.813       &  23   & $-3.2$ \\
$12_{3,10} - 11_{3,9}$   & 232544.540       &  21   & $21.0$ \\
$12_{3,9} -   11_{3,8}$   & 232546.625       &  21   & $-35.9$ \\
$12_{2,10} - 11_{2,9}$ & 232775.855       &  20   & $-7.1$ \\
$12_{1,11} - 11_{1,10}$ & 235415.394       &  20   & $24.0$ \\
$13_{1,13} - 12_{1,12}$ & 248731.752       &  20   & $-24.6$ \\
$13_{0,13} - 12_{0,12}$ & 251629.293       &  20   & $31.1$ \\
$13_{2,12} - 12_{2,11}$   & 251875.722       &  34   & $-10.0$ \\
$13_{3,10} - 12_{3,9}$   & 251927.295       &  33   & $-42.4$ \\
$13_{1,12} - 12_{1,11}$ & 255015.024       &  20   & $-26.7$ \\
$14_{1,14} - 13_{1,13}$ & 267845.608       &  37   & $-57.4$ \\
$14_{2,13} - 13_{2,12}$ & 271235.509     &  20   & $15.3$ \\
$14_{3,12} - 13_{3,11}$ & 271303.986     &  41   & $43.7$ \\
$14_{3,11} - 13_{3,10}$ & 271308.664     &  28   & $26.9$ \\
$14_{2,12} - 13_{2,11}$ & 271653.245     &  20   & $13.3$ \\

\hline
\end{tabular}

\tablefoot{
\tablefoottext{a}{Calculated with the spectroscopic constants in Table~\ref{spectro_parameter}.}}
\end{table}

\begin{table}
\begin{center}
\caption{Spectroscopic constants of HDCCC (in MHz)}
\label{spectro_parameter}
\renewcommand{\arraystretch}{1.10}
\begin{tabular}[t]{lr@{}lr}
\hline \hline
Constant\tablefootmark{a} & \multicolumn{2}{c}{This work\tablefootmark{b}}   & \multicolumn{1}{c}{Expected\tablefootmark{c}}         \\
\hline
$A$                                        & 199\,370&.(233)                      &   199\,755  \\
$B$                                        &   9\,934&.225\,82(81)             &    9\,936.8 \\
$C$                                        &   9\,450&.298\,72(81)             &  9\,452.8 \\
$D_K$                                   &       15&.95\tablefootmark{d}                              & 15.95   \\
$D_{JK} \times 10^3$         &      353&.89(50)                       &   413.19  \\
$D_J \times 10^3$              &        3&.808\,2(33)                   &   {3.79}   \\
$d_1 \times 10^6$               &   $-$256&.62(164)                  &  $-235.43$     \\
$d_2 \times 10^6$               &   $-$91&.00(250)                     &   $-107.81$    \\
$H_{KJ} \times 10^6$         &   $-$841&.(59)                         &      \\
$H_{JK} \times 10^6$         &    8&.53(122)                            &      \\
\hline
\end{tabular}\\[2pt]
\end{center}
\tablefoot{
\tablefoottext{a}{Spectroscopic constants in Watson's $S$-reduced Hamiltonian in the $I^r$ representation.}
\tablefoottext{b}{Numbers in parentheses are one standard deviation in units of the least 
significant \hfill\break digits.}
\tablefoottext{c}{Rotational constants calculated with the empirical equilibrium structure of \citet{gau99}, and vibration-rotation interaction 
constants from \citet{wu10}.   
The fourth order centrifugal distortion constants  were calculated at the B3LYP/cc-pVTZ level of theory,
and scaled by the ratio of the corresponding measured \citep{got93} and theoretical constants of H$_2$CCC.}
\tablefoottext{d}{Constrained to the theoretical value.}}
\end{table}

There is no evidence of deuterium hyperfine structure (hfs) when the fundamental transition of HDCCC is observed with high S/N at a
resolution of 1~kHz in our molecular beam, therefore no such structure should be present when this transition is observed in space.
With these measurements, lines of HDCCC of principal astronomical interest in rotational ladders with $K_a \le  3$ can now be predicted 
to an accuracy of about 0.1~km~s$^{-1}$ for transitions up to 400~GHz, allowing for precise measurements of HDCCC in the interstellar gas.

The evidence that HDCCC is the carrier of the lines observed in our laboratory discharge is overwhelming.
The close harmonicity of nine lines in the $K_a = 0$ rotational ladder with similar relative intensities confirms that  there are no misassignments.
The derived rotational constants are in excellent agreement with those predicted from a benchmark empirical equilibrium structure of 
H$_2$CCC \citep{gau99}, combined with theoretical vibration-rotation interaction constants in \citet{wu10}.  
Specifically, the measured constants $B$ and $C$ are within 0.03\% of those estimated from the theoretical structure; 
and the fourth order distortion constants are within a factor of two of the corresponding constants in H$_2$CCC, and within  20\% of those 
which were calculated here at the B3LYP/cc-pVTZ level of theory and scaled by the ratio of the corresponding experimental 
\citep{got93} and theoretical constants of  the normal isotopic species H$_2$CCC. 
Additional evidence in support of the identification is provided by the relative intensities of lines of HDCCC which are  
the same to within 30\% of those of H$_2$CCC under the same conditions, after taking into account the statistical
mixture of the deuterated acetylene precursor. 

Predictions based on this work will be available online via the Cologne Database for Molecular Spectroscopy at http://cdms.de \citep{mul05}.


\subsection{Astronomical}

The millimetre-wave observations were done in several observing sessions between September 2012 and April 2014 with the 
IRAM 30m telescope at Pico Veleta (Spain).
The EMIR receivers in the E090 and E150 configuration were employed, and the observations were made by frequency switching 
with offsets of $\pm3.9$ and $\pm7.8$~MHz in the higher frequency band. 
All four EMIR sub-bands were connected to the FTS spectrometer which was set to high resolution mode. 
The spectrum consisted of four 1.8~GHz wide sub-bands with 50~kHz channel spacing (corresponding to a velocity resolution of 
0.15~km~s$^{-1}$ at 3~mm), and a total  spectral coverage of 7.2~GHz. The $J_{K_{a}K_{c}}=5_{05}-4_{04}$ transition of H$_2$CCC in TMC-1C was observed with VESPA with a  frequency resolution of 20 KHz. 
Telescope pointing (checked about every two hours)  was accurate to $3^{\prime\prime }- 4^{\prime\prime}$.
The coordinates (see Tables~\ref{table:lineparametersC3H2} and \ref{table:lineparametersC3HD}) for TMC-1C\footnote{The source here referred to as TMC-1C, following the nomenclature from \cite{bel88}, is listed as JCMTSF J044115.2+254932 in the SIMBAD database (http://cds.u-strasbg.fr).} are from  
\citet{bel88} and \citet{ger87}, and those of L1544 coincide with  the peak of the 1.3~mm continuum dust emission from \citet{war99}.
The GILDAS\footnote{http://www.iram.fr/IRAMFR/GILDAS} software \citep{Pet05} was used for the data processing.

Observations of the fundamental rotational transition  of HDCCC near 19.4~GHz ($J_{K_{a}K_{c}}=1_{01}-0_{00}$) were made in 2007
October during a search for a rotational line of the CCCN$^{-}$ anion\footnote{The $J=2-1$ transition of CCCN$^{-}$ is  22~MHz higher in frequency than the fundamental transition of HDCCC.} in TMC-1 \citep{tha08} with the NRAO 100 m Green Bank Telescope
(GBT).\footnote{The National Radio Astronomy Observatory is a facility of the National Science Foundation operated under cooperative agreement by Associated Universities, Inc.} 
 The corresponding transition of H$_2$CCC near 20.8~GHz was also observed during the same session, but only for a fairly short time ($\sim12$~min vs. $12$~h for HDCCC).  
The observing procedure was essentially the same as that described in \citet{bru07}. The observations were done toward the ``Cyanopolyyne Peak'' of TMC-1 (see Table~5 for coordinates), and spectra were taken by position switching, at a frequency resolution of 1.5~kHz across the 50~MHz band of the GBT spectrometer
\citep[other observing conditions were summarized in][]{tha08}.

\begin{figure*}
\centering
\includegraphics [width=0.8\textwidth]{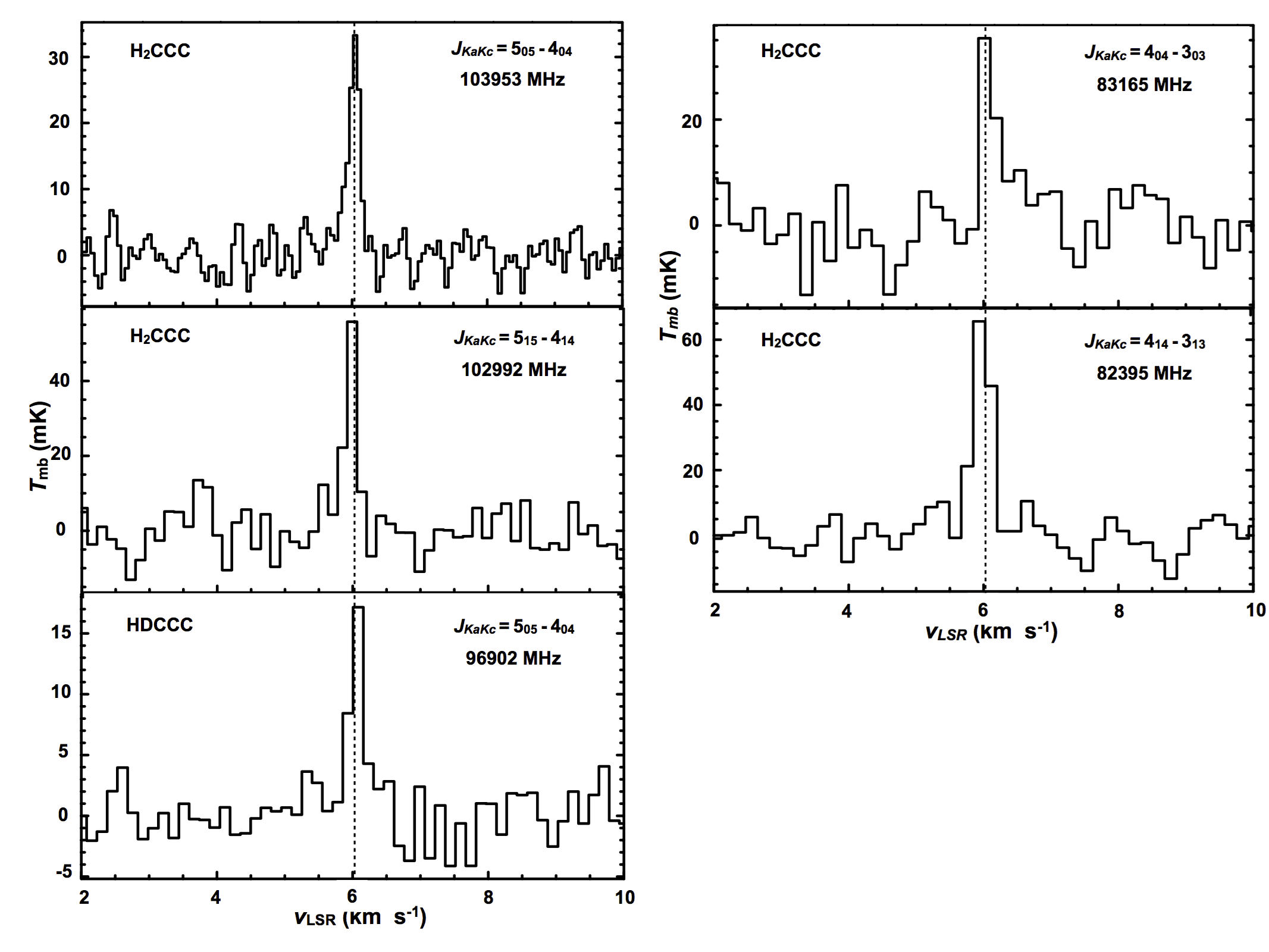}
\caption{Spectra of H$_2$CCC and HDCCC observed toward TMC-1C  with the IRAM 30~m telescope.
The integration times for H$_2$CCC were 3~h at 82 and 83~GHz, 4~h at 102 GHz and 23~h at 103~GHz. For HDCCC the integration time was 16~h.}
\label{fig:spectra_TMC1C}
\end{figure*}

\begin{figure}
\centering
\includegraphics [width=0.40\textwidth]{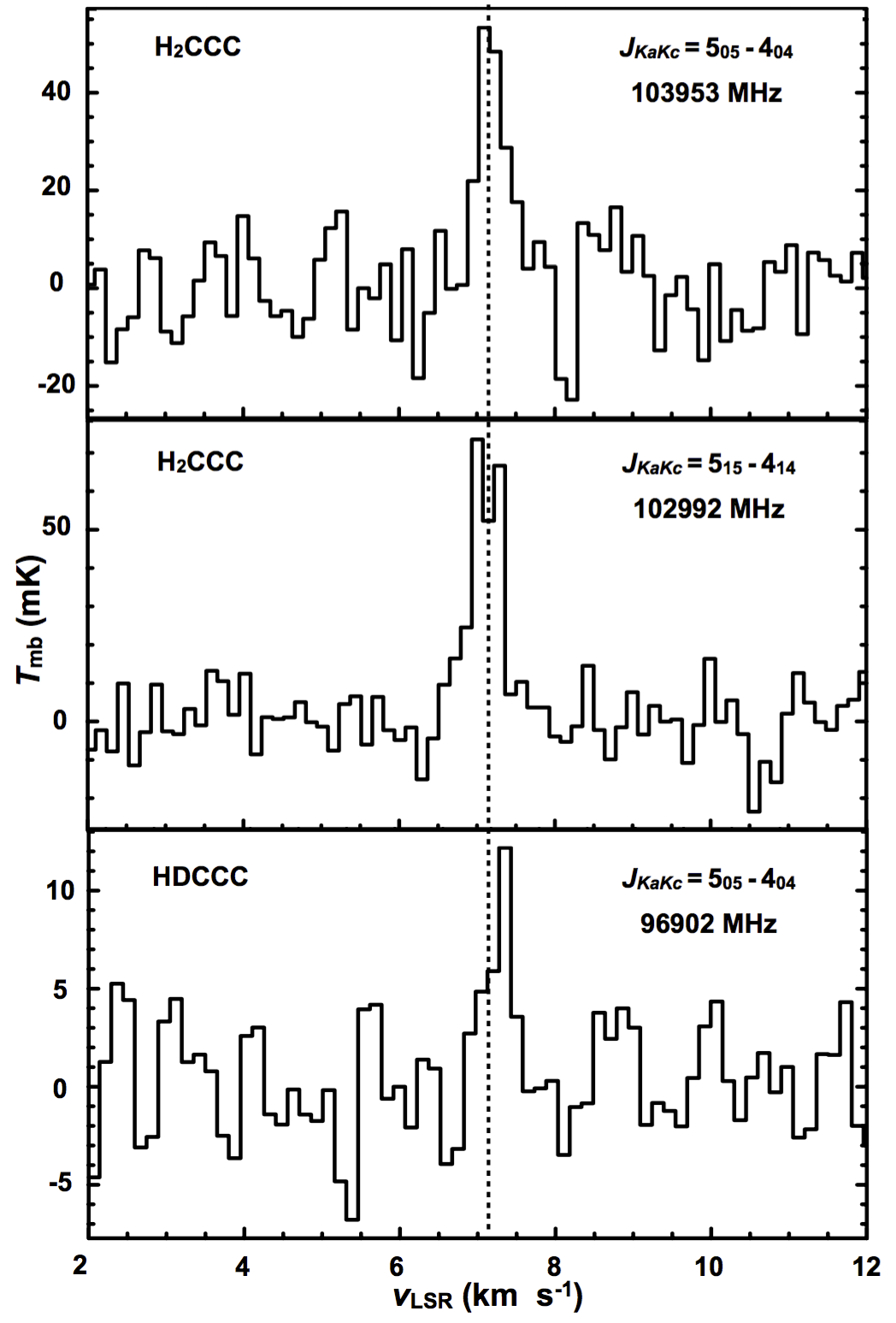}
\caption{Spectra of H$_2$CCC and HDCCC observed toward  L1544 with the IRAM 30~m telescope.
The integration times were  4~h for H$_2$CCC and 9.6~h for HDCCC.}
\label{fig:spectra_L1544}
\end{figure}

\begin{figure}
\centering
\includegraphics [width=0.35\textwidth]{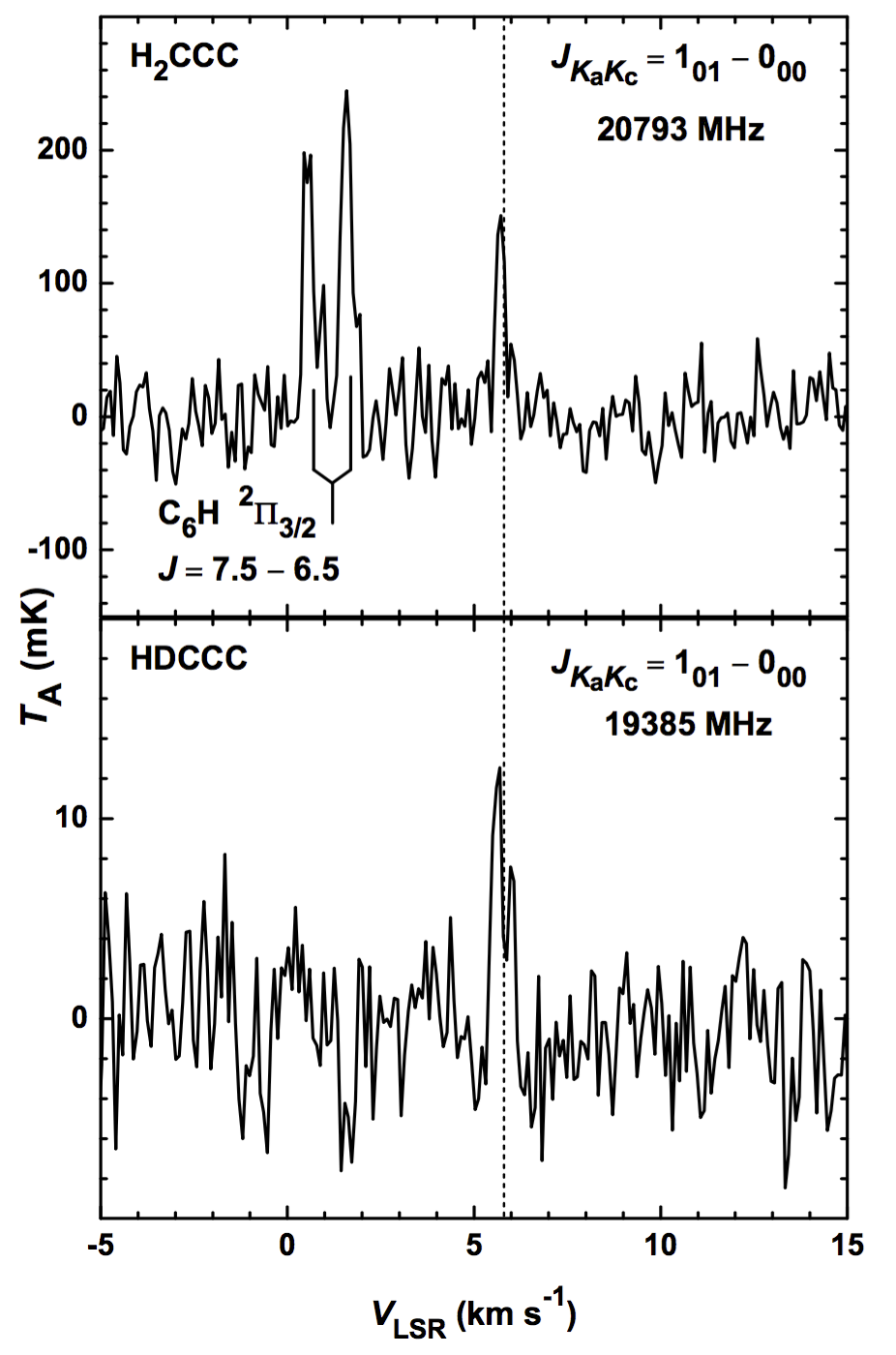}
\caption{Fundamental transitions of HDCCC and H$_2$CCC in TMC-1. Also present in the upper panel is the hyperfine-split transition of 
C$_6$H in the lower $\Lambda$ component ($e$) at 20792.872 and 20792.945~MHz. Partially resolved kinematic structure is seen in the lines of  HDCCC and C$_6$H, but not in H$_2$CCC owing to the lower signal-to-noise ratio.  The dashed line at $+5.8$~km~s$^{-1}$ 
marks the systemic velocity of TMC-1.  
The integration time was approximately 11.5~h for the HDCCC, and  12~min for the  H$_2$CCC spectrum.  Both spectra were Hanning smoothed to a frequency resolution of 6.1~kHz. A similar spectrum was observed by \citet{cer87}, see Figure 2 lower left panel in their paper, and \cite{fos01}, see Figure 1 upper panel. In the paper by \cite{cer87} the line of H$_2$CCC was reported as an unidentified feature because the laboratory spectrum of H$_2$CCC was not known yet.}
\label{fig:spectra_GBT}
\end{figure}

\begin{table*}
\caption{Observed line parameters of H$_2$CCC in TMC-1C and L1544}
\label{table:lineparametersC3H2}
\scalebox{0.95}{
\begin{tabular}{cccc ccc ccc cc}
\hline\hline
Transition                       & Frequency     &Ref.   & $E_{\rm{up}}$    & $T_{\rm{mb}}$& rms     & $W$                 & $B_{\rm{eff}}$  &$\theta_{\rm{MB}}$    
&$V_{\rm{LSR}}$\tablefootmark{a}                       &$\Delta v$\tablefootmark{a}                                       & $N$\tablefootmark{b}                           \\
 $J^{\arcmin}_{K_a K_c} - J^{\arcsec}_{K_a K_c}$            &(GHz)        &       & (cm$^{-1}$)        &(mK)                  &(mK)     & (K km~s$^{-1}$)  & (\%)              &($^{\prime\prime}$)
&(km~s$^{-1}$)                           & (km~s$^{-1}$)                                   &  ($\times 10^{10}$~cm$^{-2}$)                 \\
\hline
&&&&&\bfseries {TMC- 1C}$^{\rm c}$\\

$4_{1 4} - 3_{1 3}$ (ortho) &82.395      &1          & 6.18                     &70(9)                & 7           &0.024(2)              & 81                   &30
&5.98(1)                                             & 0.32(3)                                                 &27(1)         \\
$4_{0 4} - 3_{0 3}$ (para) & 83.165        &1         & 6.94                     &45(20)              &6           & 0.011(2)              & 81                   &30 
&6.07(2)                                             & 0.2(1)                                                   & 5.9(6)                                  \\
$5_{1 5} - 4_{1 4}$ (ortho) &102.992      &1          &9.62                     & 57(10)              &7           &0.012(2)               & 80                  &25 
&5.97(1)                                             & 0.18(3)                                                 &25(2)                         \\
$5_{0 5} - 4_{0 4}$ (para)&103.953     &1            & 10.40                  &31(5)               &3           &0.007(1)                & 80                  &25
&6.03(1)                                            & 0.21(2)                                                 &8.7(7)                       \\
&&&&&\bfseries {L1544}$^{\rm d}$\\
$5_{1 5} - 4_{1 4}$ (ortho) &102.992        &1        & 9.62                    & 69(8)                 &10         &0.037(3)              &80                   &25
&7.11(2)                                             & 0.50(5)                                                 & 83(4)      \\
$5_{0 5} - 4_{0 4}$ (para) &103.953         &1       & 10.40                  & 53(12)               &10         &0.024(4)              & 80                 &25
&7.18(3)                                             & 0.42(7)                                                 & 30(3)    \\

\hline
\end{tabular}
}
\tablefoot{
\tablefoottext{a}{Derived from a least-squares fit of Gaussian profiles to the spectra in Figure~\ref{fig:spectra_TMC1C} and~\ref{fig:spectra_L1544}.}
\tablefoottext{b}{Calculated on the assumption that $T_{ex}$ is 4~K.}
\tablefoottext{c}{Pointing position:  $\alpha _{2000}$ = 04$^h$41$^m$16$^s$.1,     $\delta _{2000}$ = +25$^\circ$49$'$43$''$.8}
\tablefoottext{d}{Pointing position:  $\alpha _{2000}$ = 05$^h$04$^m$17$^s$.21,  $\delta _{2000}$ = +25$^\circ$10$'$42$''$.8}\\
References: (1) \citet{vrt90}}

\end{table*}

\begin{table*}
\caption{Observed line parameters of HDCCC in TMC-1C and L1544}
\label{table:lineparametersC3HD}
\scalebox{1}{
\begin{tabular}{cccc ccc ccc cc}
\hline\hline

Transition                       & Frequency    &  Ref.  & $E_{\rm{up}}$    & $T_{\rm{mb}}$& rms     & $W$                 & $B_{\rm{eff}}$  &$\theta_{\rm{MB}}$    
&$V_{\rm{LSR}}$\tablefootmark{a}                       &$\Delta v$\tablefootmark{a}                                       & $N$\tablefootmark{b}                           \\
                         $J^{\arcmin}_{K_a K_c} - J^{\arcsec}_{K_a K_c}$                  &(GHz)       &           & (cm$^{-1}$)        &(mK)                  &(mK)     & (K km~s$^{-1}$)  & (\%)              &($^{\prime\prime}$)
&(km~s$^{-1}$)                           & (km~s$^{-1}$)                                   &  ($\times 10^{9}$~cm$^{-2}$)                 \\

\hline
&&&&&\bfseries {TMC- 1C}$^c$\\
$5_{0 5} - 4_{0 4}$      &96.902             &     1        & 9.70                   &17(4)                  &3           &0.005(1)            & 80                      & 27 
&6.09(2)                                       &    0.26(4)                                            & 64(7)           \\
$7_{0 7} - 6_{0 6}$      & 135.634       &        1        &18.10                 & $\le 3$               &3            &  $\cdots$       & $\cdots$           &  $\cdots$
&  $\cdots$                                   &  $\cdots$                                           &      $\cdots$      \\
&&&&&\bfseries {L1544}$^d$\\
$5_{0 5} - 4_{0 4}$      &96.902          &      1          & 9.70                   &12(4)                 &3           &0.005(1)            & 80                      & 27
&7.31(5)                                       &   0.41(1)                                             &71(7)            \\
\hline
\end{tabular}
}
\tablefoot{
\tablefoottext{a}{Derived from a least-squares fit of Gaussian profiles to the spectra in Figure~\ref{fig:spectra_TMC1C} and~\ref{fig:spectra_L1544}.}
\tablefoottext{b}{Calculated on the assumption that $T_{ex}$ is 4~K.}
\tablefoottext{c}{Pointing position:  $\alpha _{2000}$ = 04$^h$41$^m$16$^s$.1,     $\delta _{2000}$ = +25$^\circ$49$'$43$''$.8}
\tablefoottext{d}{Pointing position:  $\alpha _{2000}$ = 05$^h$04$^m$17$^s$.21,  $\delta _{2000}$ = +25$^\circ$10$'$42$''$.8}\\
References: (1) This work}
\end{table*}

\begin{table*}
\caption{Centimetre-Wave Lines of HDCCC  and H$_2$CCC in TMC-1}
\scalebox{0.95}{
\begin{tabular}{lccclccccccc}
\hline\hline

  Molecule    &    Transition
   &    Frequency   &    Ref.   &    $E_{up}$   &    $T_{\rm{A}}$\tablefootmark{a}     &   $W$ 
 &   $\eta_{\rm{MB}}$   &   $\theta_{\rm{MB}}$ &   $V_{\rm{LSR}}$\tablefootmark{a}   &   $\Delta v$\tablefootmark{a}     
&   $N$\tablefootmark{b}   \\  

 &   $J^{\arcmin}_{K_a K_c} - J^{\arcsec}_{K_a K_c}$   &   (GHz)  &      &   (cm$^{-1}$)   &   (mK)    &   (K~km~s$^{-1}$) 
&   $(\%)$      &   $(\arcsec)$ &   (km~s$^{-1}$)   &   (km~s$^{-1}$)     &   $(10^{9}\mathrm{cm}^{-2})$   \\

\hline

HDCCC         &   $1_{01}-0_{00}$      &   19.384   &   1   &   0.647   &   13(3)     &  0.006(2) &  0.880     &   37   &  5.62(3)   &   0.28(6)      &  118(36)        \\

                        &                                       &                              &        &                 &   10(6)   &&&    & 6.02(2)    &   0.14(11) &                                        \\

H$_2$CCC ($para$)  & $1_{01}-0_{00}$        & 20.792     &   2   &   0.694   &   158(19) &  0.054(18)  &  0.867    &  36  & 5.71(7)   &   0.28(4)      & 769(231)\tablefootmark{c}   \\

\hline
\end{tabular}
}
\tablefoot{Pointing position for TMC-1: $\alpha_{2000}=04^{\mathrm{h}}41^{\mathrm{m}}42.49^{\mathrm{s}}, 
\delta_{2000}=+25^{\circ}41\arcmin26.9\arcsec$.
Estimated $1\sigma$ uncertainties (in parentheses) are in units of the least significant digits.
\tablefoottext{a}{Derived from a least-squares fit of Gaussian profiles to the spectra in Figure~\ref{fig:spectra_GBT}.}
\tablefoottext{b}{Calculated on the assumption that  $T_{ex}=4$~K, the value derived for H$_2$CCC in TMC-1 by \citet{kaw91}.}
\tablefoottext{c}{For {\sl para} H$_2$CCC.  On the assumption that the {\sl ortho}/{\sl para} ratio is 3, $N(ortho+para)=3076(923)\times10^{9}$~cm$^{-2}$.} 
References: (1) This work; (2) \citet{vrt90}}
\label{table:TMC-1}
\end{table*}

\subsubsection{HDCCC}
\label{subsec:$l$-C$_3$HD}

The $5_{0,5} - 4_{0,4}$ transition of HDCCC at  96.9~GHz was observed in TMC-1C and L1544 (Figures~\ref{fig:spectra_TMC1C}  
and~\ref{fig:spectra_L1544}). 
Because the $V_{\rm{LSR}}$  is in excellent agreement with that of other molecules in these two narrow line sources, these spectra provide  strong evidence that HDCCC has been observed for the first time in the interstellar gas.
The probability of a misidentification due to a line of another carrier coinciding to within twice the typical line width of about 200~kHz 
is very small ($\sim 2 \times 10^{-3}$), owing to the very low  density of lines in these two sources of about 20 lines in each 1.8~GHz 
wide sub-band.
Additional evidence in support of the astronomical identification of HDCCC was sought by means of other transitions in the millimetre band, however the rotational spectrum of HDCCC is not rich in lines.
On the assumption that only levels in the $K_a = 0$ ladder are appreciably populated in these cold dense cores 
(those with $K_a \ne 0$  are not metastable and are higher in energy by $\ge 12$~K), there are three transitions that are readily accessible with present ground based telescopes which might be detectable with deep integrations: the two lowest ones at 19 and 38~GHz, and one at 
136~GHz.


Our initial attempt to detect a second line of HDCCC in the millimetre band was inconclusive.
Although the transition at 135.6~GHz was not detected in TMC-1C with the IRAM 30~m telescope, our upper limit of the main beam 
temperature at this frequency ($\le 3$~mK) is greater than that predicted from the  column density derived from the line at 96.9~GHz on 
the assumption that the excitation temperature is 4~K (see Section~\ref{subsec:Analysis}).  
We then realized that the $1_{0,1} - 0_{0,0}$ transition at 19~GHz  was covered in October~2007 in a deep search for the 
CCCN$^-$ anion in TMC-1 with the GBT \citep{tha08}.
Shown in Figure~\ref{fig:spectra_GBT} are the spectra with the fundamental transition of HDCCC and H$_2$CCC in TMC-1 at
the precise $V_{\rm{LSR}}$ of $5.8 \pm 0.1$~km~s$^{-1}$ in this much studied position in the Taurus molecular cloud, 
confirming our initial assignment of the line at 96.9~GHz to the $5_{0,5} - 4_{0,4}$ transition of HDCCC.
As in the laboratory spectrum (Section~\ref{sec:Laboratory}), there is no evidence of deuterium hyperfine structure in
the line in TMC-1 at 19~GHz.


\subsubsection{H$_2$CCC}

Simultaneous with the observations of HDCCC (Section~\ref{subsec:$l$-C$_3$HD}), a {\sl para}  ($5_{0,5} - 4_{0,4}$) and an 
{\sl ortho} (5$_{1,5}$ - $4_{1,4}$) transition of the main isotopic species H$_2$CCC were observed in TMC-1C and L1544 in 
September 2012; and two additional transitions ($4_{0,4} - 3_{0,3}$ and $4_{1,4} - 3_{1,3}$)  were observed in 
TMC-1C in 2014.
The line parameters and derived column densities are reported in Table~\ref{table:lineparametersC3H2},
and the spectra are shown in Figures~\ref{fig:spectra_TMC1C} and~\ref{fig:spectra_L1544}.

Owing to the two equivalent off-axis hydrogen atoms, H$_2$CCC has {\sl ortho} and {\sl para} symmetry states with a
relative statistical weight ({\sl ortho/para}) of 3:1, where rotational levels with odd $K_a$ have {\sl ortho} symmetry
and those with even $K_a$ have {\sl para} symmetry.
Monodeuterated HDCCC does not have {\sl ortho/para} symmetry.
It has been shown that the {\sl ortho/para} ratio might depart from the statistical value, especially at the low temperatures which characterize cold dark cores such as TMC-1C and L1544  \citep[$T = 4-10$~K,][]{park06}.
While analysing our data we do not constrain the value of the {\sl ortho/para} ratio, except in the case of TMC-1 where just one ($para$) line has been observed. In this case the {\sl ortho/para} ratio was constrained to the statistical value (see Section~\ref{subsec:Analysis}).


\begin{table}
\caption{Column density and Excitation Temperature of {\sl ortho} H$_2$CCC in TMC-1C}
\center
\begin{tabular}{lccc}
\hline\hline
\multicolumn{1}{l}{Property} 
& \multicolumn{1}{c}{LTE} 
& \multicolumn{1}{c}{}
&\multicolumn{1}{c}{RADEX}    \\
\hline

$N$(cm$^{-2}$)    & $2.6 \times 10^{11}$\tablefootmark{a}           &          & $1.8 \times 10^{11}$      \\

$T_{\rm{ex}}$~(K)                                  &  4\tablefootmark{b}                                     &         &  4.6\tablefootmark{c}     \\      

$T_{\rm{kinetic}}$~(K)                           &  $\cdots$                            &         &  10.0                                 \\                            

\hline
\end{tabular}
\tablefoot{
\tablefoottext{a}{Average value of the column densities reported in Table~\ref{table:lineparametersC3H2} for $ortho$-H$_2$CCC in TMC-1C}
\tablefoottext{b}{$T_{\rm{ex}}$ assumed by analogy with \citet{cer91} and \citet{kaw91}.}
\tablefoottext{c}{Approximate mean of 4.4~K for the $5_{1,5}  - 4_{1,4}$ transition and 4.8~K for
the $4_{1,4}  - 3_{1,3}$ transition at $\log n = 4.25$ and $\log N = 11.25$.}}

\label{Comparison}
\end{table}

\begin{table*}
\caption{Isomeric and D/H Ratios of H$_2$CCC and $c$-C$_3$H$_2$ in Six Galactic Regions }
\scalebox{0.85}{
\begin{tabular}{lccccccccc}
\hline \hline
\multicolumn{1}{l}{Property} 
&\multicolumn{2}{c}{Prestellar Core}
&\multicolumn{2}{c}{Protostellar Core}
& \multicolumn{1}{c}{Dark Cloud} 
&\multicolumn{1}{c}{Translucent\tablefootmark{a} }
&\multicolumn{1}{c}{Diffuse\tablefootmark{b}}
&\multicolumn{1}{c}{PDR\tablefootmark{c}}   \\

\cline{2-3}
\cline{4-5}

\multicolumn{1}{c}{}  
&   \multicolumn{1}{c}{TMC-1C}  
&   \multicolumn{1}{c}{L1544}  
&   \multicolumn{1}{c}{L1527\tablefootmark{d}} 
&   \multicolumn{1}{c}{Cha-MMS1\tablefootmark{e}} 
&  \multicolumn{1}{c}{TMC-1}
&   \multicolumn{1}{c}{(Average)}    
&  \multicolumn{1}{l}{(Mean)}
&  \multicolumn{1}{l}{}
&  \multicolumn{1}{l}{}\\

$N$($c$-C$_3$H$_2$)~cm$^{-2}$ &  ${2.2\tablefootmark{h} \times 10^{13}}$   & ${3.7\tablefootmark{h}  \times 10^{13}}$  &  $1.3\times10^{13}$ &  $2.7\times10^{13}$  &  $0.58^{\rm{g}}\times 10^{14}$      &   $2.4 \times 10^{13}$                                                                       &    $3.3 \times 10^{12}$                   &   $9.3 \times 10^{12}$                                   \\

$N$(H$_2$CCC)~cm$^{-2}$       &   $\mathbf{3.3 \times 10^{11}}$    &  $\mathbf{1.1 \times 10^{12}}$  &  $1.1\times10^{12}$&  $\cdots$   &  $(2.1^{\rm{g}} \text{-} \mathbf{3.1}) \times 10^{12}$            &   $2.5 \times 10^{12}$                                                                &   $1.9 \times 10^{11}$                    &   $2.7 \times 10^{12}$                             \\

\multicolumn{8}{l}{}                                    \\

$c$-C$_3$H$_2$/H$_2$CCC        &       {\bf 67\bf$\pm \bf7$}\tablefootmark{f}                     &    {\bf 32\bf$\pm \bf4$}        &   $12$  &  $\cdots$   &  28$ \pm$6\tablefootmark{g}  
&   9       &  17   &     3          \\

$c$-C$_3$HD/$c$-C$_3$H$_2$       & $(5\text{-}13)\%$\tablefootmark{h} 10\%$^{\rm{i}}$       &  $(12 \text{-} 17)\%$\tablefootmark{h} 15\%$^{\rm{i,j}}$  &  $(7\text{-}18)\%$ &  $(5 \text{-} 34)\%$  
                                                     &   8\%\tablefootmark{i}  5\%\tablefootmark{k}     &    $\cdots$                   &            $\cdots$                     &             $\cdots$                       \\  

$c$-C$_3$D$_2$/$c$-C$_3$HD       & $(3\text{-}15)\%$\tablefootmark{h}                            &  $(7 \text{-} 17)\%$\tablefootmark{h}   &  $2.6\%$\tablefootmark{l}   &  $\cdots$ 
                                                   &   $\cdots$     &    $\cdots$                   &            $\cdots$                     &             $\cdots$                       \\  
                                                                                                                                                                                                                                                                                           
HDCCC/H$_2$CCC                   &    $\mathbf{19 \pm 5}$\%                    &     $\mathbf{6 \pm 1.3}$\%   &  $\cdots$  &  $\cdots$  &     $ \mathbf{4 \pm 1.6}$\%              &  $\cdots$    &      $\cdots$           &    $\cdots$            \\

\hline
    \end{tabular}
    }

    \tablefoot{
Quantities in {\bf boldface} are from this work.
\tablefoottext{a}{From Table~20 in \citet{tur00}, and references therein.}
\tablefoottext{b}{\citet{lis12}.}
\tablefoottext{c}{\citet{pet12}.}
\tablefoottext{d}{From Table~1 in \citet{Sak13}, and references therein.}
\tablefoottext{e}{\citet{cor12}.}
\tablefoottext{f}{See Sections~\ref{subsec:Excitation} and \ref{sec:Discussion}.}
\tablefoottext{g}{\citet{fos01}.}
\tablefoottext{h}{\citet{spe13}.}
\tablefoottext{i}{\citet{bel88}.}
\tablefoottext{j}{\citet{ger87}.}
\tablefoottext{k}{ See Table~12 in \citet{tur01}.}
\tablefoottext{l}{\citet{tok13}.}}

\label{D/Hratios}
\end{table*}

\section{Analysis}
\label{subsec:Analysis}

\begin{figure}[]
\includegraphics [width=0.47\textwidth]{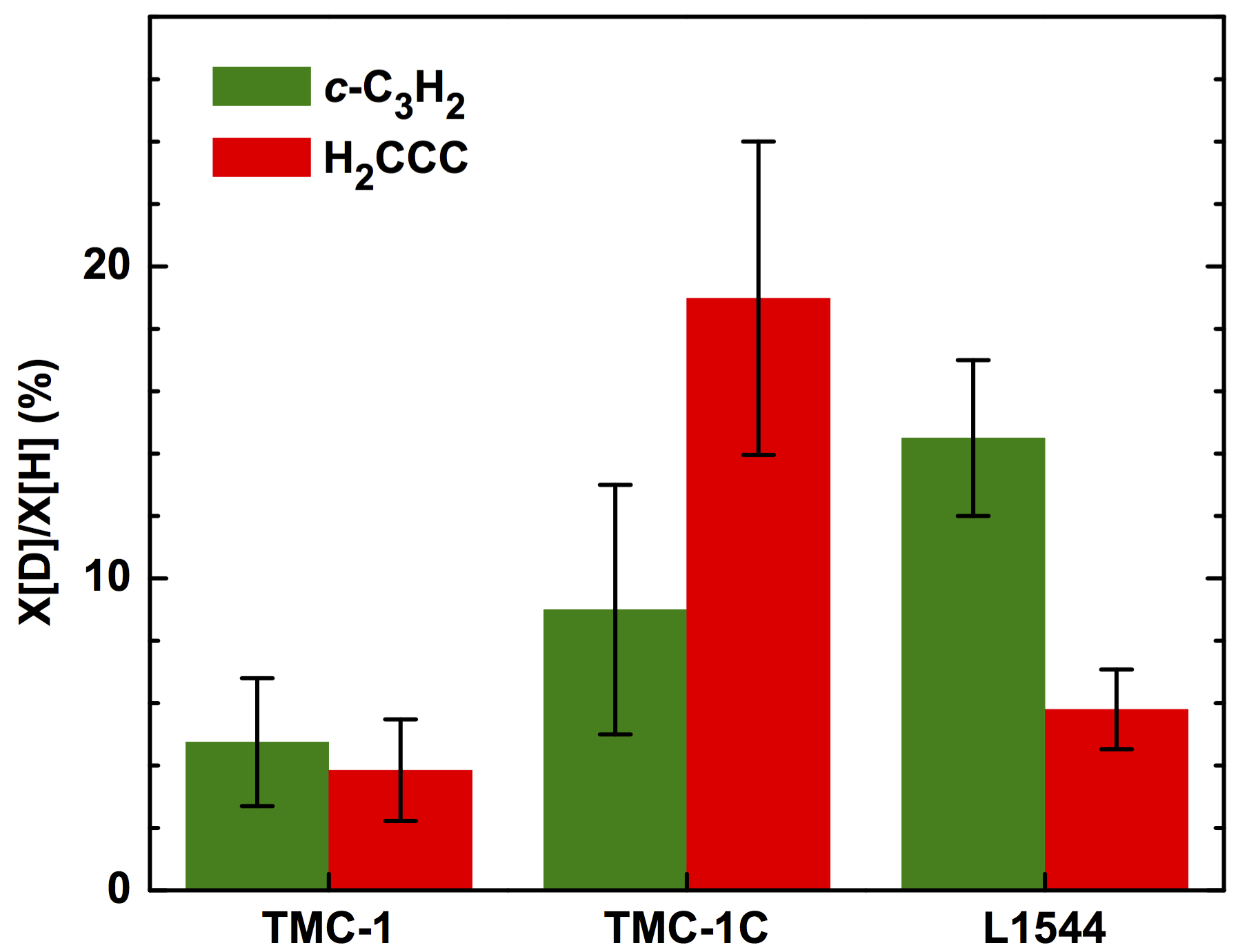}

\caption{D/H isotopic ratio in the carbene isomers H$_2$CCC and $c$-C$_3$H$_2$ in three dense cores in the Taurus molecular cloud:
TMC-1, TMC-1C, and L1544.
For $c$-C$_3$H$_2$, the heights of the bars indicate the average value, and the error bars the range of values inferred in previous studies of
TMC-1 \citep{tur01}, and of TMC-1C and L1544 \citep{spe13}.
}
\label{fig:D-H(Ratio)}
\end{figure}

The column densities in Tables~\ref{table:lineparametersC3H2}, \ref{table:lineparametersC3HD}, and \ref{table:TMC-1} were calculated with the following expression for optically thin transitions in rotational equilibrium at a temperature $T_{\mathrm{ex}}$ \citep{gol99}: 

\begin{equation}
\frac{N_{u}}{g_{u}} =\frac{J(T_{\mathrm{ex}})}{J(T_{\mathrm{ex}})-J(T_{\mathrm{bg}})}\frac{3kW}{8\pi^{3} \nu S \mu^{2}} = \frac{N}{Z} 
e^{\frac{E_u}{kT_{\mathrm{ex}}}}~,
\end{equation}

\noindent where $N_u$, $g_u$, and $E_u$ are the column density, degeneracy, and energy of the upper state of the transition; $W$ is the  integrated intensity, $\nu$ the frequency, $S$ the rotational line strength, $\mu$ the dipole moment, $Z$ the rotational partition function, and 
$J(T) \equiv {\frac{h\nu}{k}}(e^{\frac{h\nu}{kT}}-1)^{-1}$. The factor $J(T_{\mathrm{ex}})/[J(T_{\mathrm{ex}})-J(T_{\mathrm{bg}})]$ accounts for line absorption of the cosmic background radiation ($T_{\mathrm{bg}}=2.7$~K), and is significantly larger than unity at the low excitation temperatures ($T_{\mathrm{ex}} < 10$~K) inferred for 
H$_2$CCC and other polar molecules in cold dark clouds. The column densities were calculated assuming an excitation temperature of 4 K, by analogy with earlier work from \citet{cer91} and \citet{kaw91}. On this assumption, from the column densities derived from the {\sl para} and {\sl ortho} lines of H$_2$CCC in TMC-1C and L1544, we infer a {\sl ortho/para} ratio of $\sim$3, which is consistent with the canonical ratio of 3.
There are three principal sources of uncertainty in our derived column densities of H$_2$CCC and HDCCC:  the excitation temperature,  the
{\sl ortho/para} ratio in H$_2$CCC, and possible systematic uncertainties in the integrated areas of the observed line profiles.
The results here allow us to derive for the first time the D/H ratio in H$_2$CCC in three sources. 
On the assumption that $T_{ex}$ is 4~K in both H$_2$CCC and HDCCC, the D/H ratio in H$_2$CCC is $19 \pm 5$\% 
in TMC-1C and $6 \pm 1.3$\% in L1544 (Figure~\ref{fig:D-H(Ratio)}).  
However, more observations are needed to better constrain the excitation temperature in both isotopic
species, and in turn the D/H ratio.
Similarly, the centimetre-wave observation of the fundamental transition of HDCCC and {\sl para}  H$_2$CCC allowed us to estimate 
an approximate D/H ratio in TMC-1.
We find that the extent of the deuteration in TMC-1 ($4 \pm 1.6$\%) is comparable to that in L1544,  but is two to five times smaller than 
in TMC-1C. 

\vspace{0.25cm}
\subsection{Excitation and abundance of H$_2$CCC and HDCCC}
\label{subsec:Excitation}

To determine whether LTE is a good approximation for estimating the rotational excitation temperature ($T_{\rm{ex}}$) and 
column density ($N$) of H$_2$CCC and HDCCC (Section~\ref{subsec:Analysis}), 
$T_{\rm{ex}}$ and $N$ were compared with the corresponding properties of H$_2$CCC obtained from statistical equilibrium calculations with RADEX \citep{vdt07}.
The calculations were done for a uniform spherical geometry for TMC-1C, a line width of 0.3~km~s$^{-1}$ (FWHM), 
a kinetic temperature of 10~K, and a radiation temperature of 2.73~K.
Following \citet{cer99}, we used collisional de-excitation rates that are two times higher than those for {\sl ortho}-H$_2$CO with 
{\sl para}~H$_2$ as provided in the LAMDA database \citep{sch05,wie13}.
Only  the lowest 18~rotational levels in the $K_a = 1$ ladder of H$_2$CCC were included, because the cross-ladder collisional rates 
($\Delta K = 2$) are nearly an order of magnitude lower.
Similarly, collisions with {\sl ortho}-H$_2$ were neglected, because the {\sl ortho/para} ratio for H$_2$ is negligible at 10~K \citep[cf.,][]{tro09}.

\begin{figure}
\includegraphics[scale=0.85]{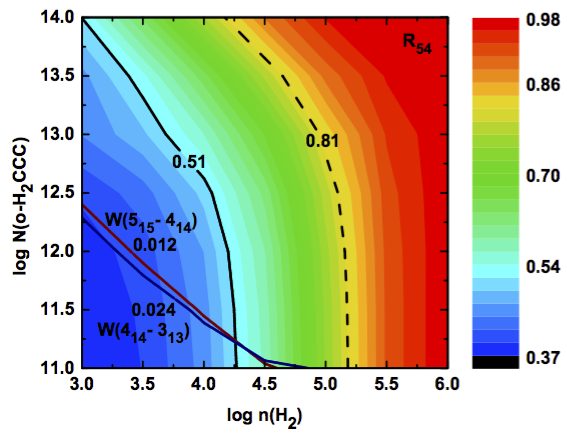}
\caption{Integrated area (W) for the $4_{1,4} - 3_{1,3}$ transition
(blue contour) and  $5_{1,5} - 4_{1,4}$ transition 
(red contour)  of H$_2$CCC, and the ratio of the integrated areas R$_{54}$ (solid black contour) 
versus the total column density of {\sl ortho} H$_2$CCC and the H$_2$ density.   }
\label{fig-WRatio}
\end{figure}

\begin{figure}

\includegraphics[scale=0.85]{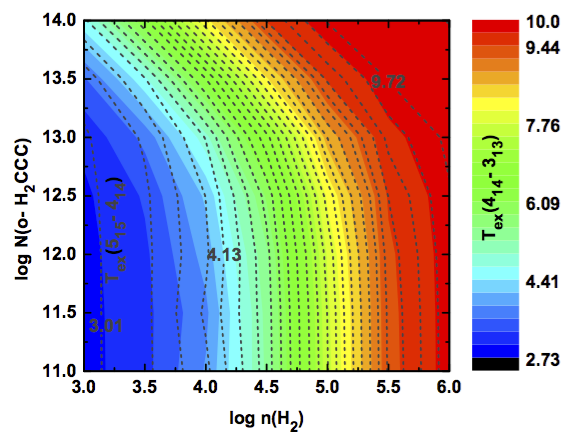}
\caption{Dependence of the excitation temperature ($T_{\rm{ex}}$) of the $4_{1,4} - 3_{1,3}$ and $5_{1,5} - 4_{1,4}$ transitions of 
{\sl ortho} H$_2$CCC on the column density and the H$_2$ density, on the assumption that the kinetic temperature is 10~K. 
The color scale applies to the $4_{1,4} - 3_{1,3}$ transition and the dashed contours to the $5_{1,5} - 4_{1,4}$ transition.}
\label{fig-Tex}
\end{figure}

\begin{figure}

\includegraphics[scale=1.2]{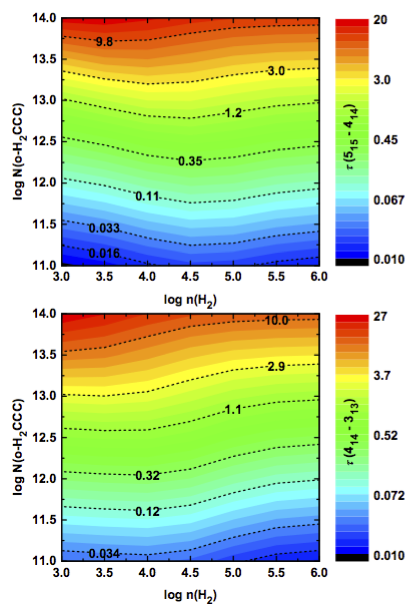}
\caption{Dependence of the optical depths ($\tau$) of the $4_{1,4} - 3_{1,3}$ and $5_{1,5} - 4_{1,4}$ transitions of {\sl ortho} H$_2$CCC 
on the column density and the H$_2$ density.   
}
\label{fig-Tau}
\end{figure}

Summarized in Figures~\ref{fig-WRatio} - \ref{fig-Tau} are the results of the statistical equilibrium calculations.
It is evident that the integrated areas (W) of the two {\sl ortho} transitions of H$_2$CCC and the ratio of the two, imply that 
$\log n({\rm{H_2}}) \sim 4.25$ and $\log N \sim 11.25$ 
(i.e., $n({\rm{H_2}}) = 2 \times 10^4~{\rm{cm}^{-3}}$ and $N = 2 \times 10^{11}$~cm$^{-2}$; Figure~\ref{fig-WRatio}).
Under the same conditions we estimate that $T_{\rm{ex}} \sim 4$~K (Figure~\ref{fig-Tex}), and H$_2$CCC is optically thin in both 
transitions (i.e., $\tau \le 0.05$ in $4_{1,4} - 3_{1,3}$ and  $5_{1,5} - 4_{1,4}$; Figure~\ref{fig-Tau}).
As a result, we find that the rotational temperature derived for H$_2$CCC in TMC-1 by \citet{cer91} and \citet{kaw91} is valid also for TMC-1C, and hence a valid assumption also for L1544. We also find that the column density derived for the $ortho$ H$_2$CCC on the assumption of LTE is essentially the same as that derived with RADEX (see Table~\ref{Comparison}).

\section{Discussion}
\label{sec:Discussion}

The present work is a preliminary study of different isomers and  isotopologues of the C$_3$H$_2$ system. Isomers and isotopologues are precious tools for astrochemists as they convey useful information to disentangle different chemical routes which might correspond to different physical conditions. In particular, the astrophysically relevant properties of the C$_3$H$_2$ system studied in this paper are the D/H abundance ratio in each isomer, and the cyclic to linear abundance ratio.

The cyclic to linear isomeric ratio of C$_3$H$_2$ appears to depend on the region where the isomers are observed, namely the $c$-C$_3$H$_2$/H$_2$CCC ratio increases with decreasing electron abundance \citep{fos01}. A summary of the $c$-C$_3$H$_2$/H$_2$CCC ratio in different environments is reported in Table 7. In dense molecular clouds the ratio is between 20 and 40 \citep{kaw91,cer91,fos01}, while in diffuse clouds it is lower by one order of magnitude \citep{cer99}. In the Horsehead Nebula PDR the $c$-C$_3$H$_2$/H$_2$CCC ratio has a value of 3-5 in the diffuse gas, and it increases by a factor of 4 when penetrating in the denser region of the cloud \citep{tey05}. Towards the Orion Bar a $c$-C$_3$H$_2$/H$_2$CCC ratio of 34 has been observed very recently \citep{cua15}. Perhaps alternative formation/destruction routes happen towards the Orion Bar, which is a quite extreme environment ($>$300 times the FUV radiation flux with respect to the Horsehead Nebula), that might allow endothermic reactions and reactions with high energy barrier to become efficient. The $c$-C$_3$H$_2$/H$_2$CCC ratio in TMC-1C is about two times higher than that in TMC-1, along the filament in TMC-1 (see Figure 3 in \citet{fos01}, and in L1544. The causes for this behaviour might be a systematic error in the H$_2$CCC column density (see Section \ref{subsec:Analysis}), the overall difficulty in deriving accurate column densities of $c$-C$_3$H$_2$, or different physical conditions in TMC-1C with respect to TMC-1 and L1544.

The variation of the cyclic to linear ratio in different environments is related to distinct destruction and possibly formation paths for the two isomers. These paths are still to be fully understood. In \cite{tal09} are presented the results of calculations on the dissociative recombination of cyclic and linear C$_3$H$_3^+$ with electrons. It is shown that the formation of $c$-C$_3$H$_2$ from $c$-C$_3$H$_3^+$ is more efficient than the formation of H$_2$CCC from $l$-C$_3$H$_3^+$. A similar result was obtained by \cite{ada05}, they in fact observed in an afterglow experiment that the cyclic C$_3$H$_3^+$ recombines with electrons faster than the linear isomer. Interestingly, \cite{cha13} show that using their new semiempirical model to calculate branching ratios, while $c$-C$_3$H is the main product of the dissociative recombination of $c$-C$_3$H$_2^+$ with electrons, $l$-C$_3$H is not the main product of the recombination of $l$-C$_3$H$_2^+$ with electrons. Unfortunately the branching ratios of the dissociative recombination of linear and cyclic C$_3$H$_3^+$ are not provided in \cite{cha13}.

The formation of $c$-C$_3$HD and $c$-C$_3$D$_2$ in dense cores has been long studied \citep{ger87, bel88, spe13}, and it is believed to happen through subsequent deuteration of $c$-C$_3$H$_2$ via reactions with H$_2$D$^+$, D$_2$H$^+$ and D$_3$$^+$ followed by the dissociative recombination of the ionic intermediate with electrons (see Figure 3 in \citet{spe13}). The intermediates of this reaction scheme are C$_3$H$_3^+$, C$_3$H$_2$D$^+$, C$_3$HD$_2^+$, and C$_3$D$_3^+$. Shown schematically, singly deuterated carbenes are produced by the formal reactions 
\begin{align}\label{eq:1}
&\mathrm{RH_2 + H_2D^+ \rightarrow RH_2D^+ + H_2} \\
&\mathrm{RH_2D^+ + e^- \rightarrow RHD +H}
\end{align}
where RH$_2$ $\equiv$ $c$-C$_3$H$_2$ (or H$_2$CCC), and the doubly deuterated carbenes by the reactions
\begin{align}
&\mathrm{RHD + H_2D^+ \rightarrow RHD_2^+ + H_2} \\
&\mathrm{RHD_2^+ + e^- \rightarrow RD_2 +H}\label{eq:2}
\end{align}

For simplicity just the reactions with H$_2$D$^+$ are shown, but the same set of reactions will proceed with D$_2$H$^+$ and D$_3^+$.
In Table 7 and Figure 4 are reported the D/H abundance ratios in both the cyclic and linear isomers of C$_3$H$_2$ in TMC-1C, L1544 and TMC-1. The D/H abundance ratio in both the cyclic and linear form of C$_3$H$_2$ are similar in all our observations, suggesting that the deuteration of the linear isomer might follow the same reaction scheme as the cyclic, with the difference that the deuteration of the linear isomer will not proceed as straightforward as the cyclic. In contrast with the cyclic, the linear deuterated intermediate ion (H$_2$CCCD$^+$) will have to undergo atom exchange or structural rearrangement while recombining with electrons, otherwise it will react back to H$_2$CCC. More experimental and theoretical studies are required to understand the detailed mechanism of deuteration of H$_2$CCC.
TMC-1C shows an enhanced deuteration in the linear isomer compared to the cyclic. Given the difficulty in deriving accurate column densities for $c$-C$_3$H$_2$, and the possibility to have a systematic error in the column density derived for H$_2$CCC in TMC-1C, the authors do not feel the necessity to put too much emphasis on this result.

In dense prestellar and protostellar dark cloud cores,
the c-C$_3$HD/c-C$_3$H$_2$ and c-C$_3$D$_2$/c-C$_3$HD ratios are very similar to those in dark clouds (Table~\ref{D/Hratios}), 
in spite of the different mechanisms thought to govern the abundances of unsaturated carbon chains and carbenes in these regions.  
In addition to the ion-molecule processes that produce carbon chains and carbenes in prestellar cores and cold dark clouds, sublimation of methane from grain mantles warmed up by the faint protostar is hypothesized to yield elevated abundances of carbon chains and carbenes in low-mass protostellar cores \citep[see][for a summary of the proposed mechanisms]{cor12}.  
Similar D/H ratios in prestellar and protostellar cores indicate that once the non-deuterated carbene is formed, the singly  and doubly deuterated carbene are produced by reaction (\ref{eq:1}) to (\ref{eq:2}), not withstanding the formation mechanism. 

A plausible explanation for the similar D/H ratios of $c$-C$_3$H$_2$ in the two types of dark cloud cores is as follows.
Owing to freeze-out in the inner regions of prestellar cores --- by analogy with the well-known case of HCO$^+$ and DCO$^+$, which are heavily depleted in the cold inner regions of prestellar cores --- carbenes reside primarily in the warmer outer regions \citep{cas02}.
Hence, if the  abundance of H$_2$D$^+$ and the electron fraction in the outer regions of prestellar cores are similar to those in protostellar cores,\footnote{This is indeed the case for low-mass cores in the Taurus complex, where \citet{cas08} report similar column densities of {\sl ortho}-H$_2$D$^+$ toward 5 prestellar and 2 protostellar cores: $N\mathrm{(o\text{-}H_{2}D^{+}) \sim1 - 5 \times 10^{13}~cm^{-2}}$.  
The electron fractions in the outer regions of typical low-mass cores as inferred from the DCO$^+$/HCO$^+$ and HCO$^+$/CO ratios
are similar: 
\noindent $\sim10^{-8}-10^{-6}$ \citep[][and references therein]{cas02}.} then according to reactions (\ref{eq:1}) to (\ref{eq:2}), 
similar c-C$_3$HD/c-C$_3$H$_2$ and c-C$_3$D$_2$/c-C$_3$HD ratios are expected.
There are no observations of HDCCC in protostellar cores, but the next larger cummulene carbene HDCCCC has been detected in L1527 at   
$\sim3-4\%$ that of H$_2$CCCC \citep{Sak09b}.
Therefore, it would not be surprising if the HDCCC/H$_2$CCC ratio in L1527 and other protostellar cores is similar to that in dense prestellar
cores and dark clouds.

In order to have a more conclusive understanding of the chemistry of the C$_3$H$_2$ system, the HDCCC/H$_2$CCC and the $c$-C$_3$HD/$c$-C$_3$D$_2$ ratios should be systematically studied in a larger sample of dense cores. For example, \citet{emp09} have investigated the use of the N$_2$D$^+$/N$_2$H$^+$ ratio as an evolutionary tracer of Class 0 protostars. A comparison with the results obtained by \citet{emp09} would clarify whether the C$_3$H$_2$ system fails to be an evolutionary tracer because carbenes are heavily depleted in the inner core of the cloud.
Future observations with interferometers, NOEMA and ALMA, will shine some light on how the deuteration of different tracers changes with radial distance from the prestellar and protostellar core.
More precise inferences derived from the observed D/H ratios in $c$-C$_3$H$_2$ and H$_2$CCC will also require further laboratory kinetic measurements. The reaction rates of the proton/deuteron transfer from H$_3^+$ (and deuterated isotopologues) to RH$_2$ (and deuterated isotopologues) have not been studied yet. Also the collisional rates for H$_2$CCC with H$_2$ are not yet available. The recombination of the linear and cyclic C$_3$H$_3^+$ with electrons has been studied by \citet{mcl05} and \cite{ada05}, but it cannot be confirmed that the skeleton of the ion is kept during the reaction.
The deuteration of the C$_3$H$_2$ isomers will be better understood if the ionic intermediates, C$_3$H$_3^+$, C$_3$H$_2$D$^+$, C$_3$HD$_2^+$, and C$_3$D$_3^+$, were observed in dark clouds. Although laboratory measured rotational spectra of these species are unavailable at present, high level quantum calculations for the C$_3$H$_3^+$ system are available to guide laboratory searches \citep{hua11}. With precise laboratory rest frequencies in hand, deep radio astronomical searches for these ions could be undertaken, and the observed column densities (or upper limits) would allow refinements to chemical models of dark clouds. 

Now that the rotational spectrum of HDCCC has been measured in the millimetre band and the molecule has been detected in three sources, a comprehensive kinetic model is needed to aid the interpretation of the existing astronomical observations and to guide future observations of the C$_3$H$_2$ system. H$_2$CCC and $c$-C$_3$H$_2$ are currently considered separately in the KIDA chemical reaction database \citep{wak15}. However, to our knowledge, no complete model that distinguishes between the possibly different deuteration pathways for these two species currently exists. With the above caveats on missing laboratory data, we are working on a chemical model that includes the deuterated forms, and spin states, of these species, following the methods laid out in \cite{sip15}. Observationally, several additional transitions should be measured in HDCCC, preferably in the centimetre band, thereby allowing a more accurate determination of the D/H ratio in H$_2$CCC. The 2$_{0,2}$-1$_{0,1}$ transition at 38.77 GHz should be six times more intense, and the two $K_a$ =1 transitions at 38.28 and 39.25 GHz should be comparable to the line at 19.38 GHz. Unfortunately it might not be feasible to observe the singly deuterated HDCCC in the less dense regions listed in Table 7 because the lines of H$_2$CCC are not very intense there. However there is a number of cold dark clouds, prestellar cores and protostellar cores in which HDCCC might be detectable, if the D/H ratio and the $c$-C$_3$H$_2$/H$_2$CCC isomeric ratio are comparable to that in TMC-1, TMC-1C and L1544. These include L1527, Lupus-1, Lupus1-A, L483, ChaMMS1, and others where H$_2$CCC has been detected \citep{Sak09b, Sak09a, cor12}. Extending the determination of the D/H ratio in both isomers might serve as an independent test of models of dark cloud chemistry.


\begin{acknowledgements}
This work has ben supported by the SFB956. S. Spezzano wishes to thank the Bonn Cologne Graduate School of Physics and Astronomy (BCGS) for financial support. Any opinions, findings, and conclusions in this article are those of the authors, and do not necessarily reflect the views of the National Science Foundation.
\end{acknowledgements}

%
%

\end{document}